\documentclass[referee]{aa}
\usepackage{graphicx,natbib}
\bibpunct{(}{)}{;}{a}{}{,}
\begin{document}

\title{Optical spectra of selected Chamaeleon I young stellar
objects\thanks{Based on observations collected at the European Southern
Observatory, Chile, (ESO proposal 67.C-00365)}}                    


\author{Carlos Saffe\inst{1}\fnmsep\thanks{On a fellowship from CONICET,
Argentina},
Mercedes G\'omez\inst{1}\fnmsep\thanks{Visiting Astronomer,
Complejo Astron\'omico el Leoncito operated under agreement between the Consejo Nacional de
Investigaciones Cient\'\i ficas y
T\'ecnicas de la Rep\'ublica Argentina and the National Universities of La
Plata, C\'ordoba, and San Juan.}, 
Sofia Randich\inst{2}, 
Diego Mardones\inst{3}, Paola Caselli\inst{2}, Paolo Persi\inst{4}, and
Germ\'an Racca\inst{1}}

\institute{Observatorio Astron\'omico de C\'ordoba, Laprida 854, 5000
C\'ordoba, Argentina \\ email: saffe@oac.uncor.edu, mercedes@oac.uncor.edu, german@oac.uncor.edu\and           
INAF--Osservatorio Astrofisico di Arcetri, L.go E. Fermi, 5 50125 Firenze,
Italia \\ email: randich@arcetri.astro.it, caselli@arcetri.astro.it \and
Departamento de Astronom\'\i a, Universidad de Chile, Casilla 36-D, Santiago, Chile \\
email: mardones@das.uchile.cl \and
Istituto Astrofisica Spaziale e Fisica Cosmica, CNR, Via del Fosso
del Cavaliere I-00133, Roma, Italia \\ email: persi@rm.iasf.cnr.it}

\offprints{C. Saffe}
                                                                                                           
\date{Received Month XX, 2003; accepted Month XX, 2003}

\abstract{We present optical spectra of eight candidate brown dwarfs
and a previously known T Tauri star (Sz 33) of the Chamaeleon I dark cloud.
We derived spectral types based on the strength of the TiO
or VO absorption bands present in the spectra of these objects as well as
on the PC3 index of \cite{mar99}. 
Photometric data from the literature are used to estimate the bolometric luminosities for these
sources.  We apply \cite{dama97}
pre-main sequence evolutionary tracks and isochrones to derive masses
and ages. Based on the presence of H$\alpha$ in emission, we confirm that
most of the candidates are young objects. Our sample however includes two
sources for which we can only provide upper limits for the emission in
H$\alpha$; whereas these two objects are most
likely foreground/background stars, higher resolution
spectra are required to confirm their true nature. Among the likely cloud
members, we detect one new sub-stellar object and three transition
stellar/sub-stellar sources.
}

\authorrunning{Saffe et al.}
\titlerunning{Cha I selected young stellar objects}

\maketitle
%

\section{Introduction}

The field of star and planet formation has attracted
much observational and theoretical interest, particularly during
the last twenty years. Impressive technological advances  
have reached high spatial resolution and sensitivity limits and
allowed us to observe true stellar embryos and their circumstellar 
environments (disks), where planets are supposed to form.
Brown dwarfs (BDs) represent an intermediate class of objects between
stars and planets. They have masses below the H-burning limit 
and then are unable to sustain stable nuclear fusion of H like
stars. However, unlike planets, BDs once formed go through a slow
gravitational contraction process that releases potential
energy and makes them shine. Thus, their brightness decreases with time as 
the energy reservoir diminishes and they cool down. Following \citet{opp00}
we adopt 80 M$_{\rm J}$ (1 M$_{\rm J}$ $=$ 10$^{-3}$ M$_{\odot}$) as the
upper mass limit for BDs and 13 M$_{\rm J}$ as the lower mass limit. Solar
metallicity objects with masses
within these limits can burn deuterium at least during part of the evolution
\citep{buw97}. Below 13 M$_{\rm J}$ there is no nuclear fusion of any element.

Several investigations suggest a star-like formation mode for BDs
\citep[see, for example,][]{el99,nate01}. 
However other two scenarios, ejection of stellar embryos \citep{recl01} and planet-like
formation in circumstellar disks \citep{pic00} have been proposed. The current
observational evidence seems to provide support to the stellar-like formation 
process \citep[see, for example,][]{mue02}, although neither the embryos
ejection nor the planet-like formation hypothesis can presently
be disregarded on observational grounds. 
Understanding the formation process
of BDs is very important as it can shed light on the mechanism of formation
of both stars and planets.

BDs are elusive because they are very faint. These objects are 
easier to detect when young since they are brighter and
indeed several young BDs are known to date
in nearby dark clouds \citep[see, for example,][]{bri98,com00,bri02}.
A complete census of these objects in a given star-forming region would allow us to
carry out a meaningful estimate of their real number and thus a measurement
of the IMF of the cloud. However a considerable observational effort
is required to fully accomplish this aim. 

In this contribution we present results of optical 
spectroscopic observations of eight candidate young BDs in the Chamaeleon I 
dark cloud proposed by several recent infrared surveys aimed 
at finding sub-stellar objects in this star-forming region
\citep[see, for example,][]{per00,goke01}. An optical spectrum of Sz 33, 
a previously known T Tauri star of the cloud, is also presented. 
Our final goal is to carry out a precise determination of the complete
IMF of the cloud. 

In \S 2 we describe the observations and data reduction.  In \S 3 we present
our analysis and results. We derive spectral types from the TiO and VO
indexes and the PC3 index of \cite{mar99} which combined with published
photometry allows us to place the stars in the HR diagram. We conclude with
a brief summary in \S 4.    


\section{Observations and data reduction}

The observations were carried out from May to July 2001
in service mode with FORS1 on the ESO-VLT UT1. This instrument
has different configurations\footnote{See the FORS1+2 User Manual for a full description
on the different observing modes of this instrument, available at:
http://www.eso.org/instruments/fors1/.}.
We used the MOS (multi-object spectroscopy) sub-mode in combination with
the grisms GRISM$-$600R and GRISM$-$600I 
to cover the spectral range between 5300 and 9000 \AA, approximately, on a
2048$\times$2048 pixels CCD detector. 
These grisms provide resolving powers R$=1230$ and 1530, respectively.
The corresponding spectral dispersions are 1.08 and 1.06 \AA/pix. 
Two order separation filters, GG435+31 and OG590+72, were employed
with grisms GRISM$-$600R and GRISM$-$600I, respectively, to isolate
the useful spectral range for each grism. 

MOS has 19 movable slitlets that can be
displaced by linear guides to any position along the dispersion
direction in the field of view. In addition, the instrument
can be rotated around its optical axis allowing a wide variety 
of target configurations.  The width of each individual slitlet is
also adjustable. For our observations we selected a width of 1$''$ for all 
the 19 slitlets. Even-numbered slitlets are 20$''$ long and odd-numbered
are 22$''$ long. 

In addition to the VLT observations we obtained spectra of two relatively
bright objects (ISO-ChaI 126 and ISO-ChaI 237, see Table 2) with the 2.15-m
telescope at the CASLEO (San Juan, Argentina) in April 2000. We used the 
REOSC spectrograph in SD (simple dispersion) mode with a 300 l/mm grating
blazed at 5000 \AA~ and a long slit opened to $\sim$ 2$''$. We registered the
spectral range between 4500 and 8500 \AA~ on a 1024$\times$1024 thinned TEK CCD detector
at a dispersion of $\sim$ 4 \AA/pix, resulting in a resolving power R$=750$.

We reduced the data using standard IRAF procedures\footnote{IRAF is distributed
by the National Optical Astronomy Observatory, which is operated by
the Association of Universities for Research in Astronomy, Inc.
under contract to the National Science Foundation.}. We trimmed the CCD frames
at each end of the slit, corrected for the bias level and divided by
appropriate normalized flat-fields. The spectra were extracted using 
the NOAO task {\it apextract} with an aperture of 5 pixel radius.
A sky subtraction was
done by fitting a polynomial to the regions on either
side of the aperture. Residual night sky lines were removed by
interpolating across
them.  A non linear low order fit to the lines in the
HeNeAr lamp (VLT spectra) or CuNeAr lamp (CASLEO data) was used to wavelength
calibrate the spectra. Typical RMS for the wavelength solutions are 0.07 and
0.04 for the grisms GRISM$-$600R and GRISM$-$600I, respectively. 
We obtained values of $\sim$ 0.5 for the RMS corresponding to the CASLEO
solution.  The VLT-FORS spectra have been flux-calibrated and
normalized to the flux at 6800 \AA. CASLEO data are presented in count
units as no flux standards were observed in this case. 

\begin{table}
\caption[]{Observed Objects} \label{LOGS}
\begin{tabular}{lllrrcl}
Name & $\alpha$(2000.0) & $\delta$(2000.0)  & IT R  & IT I  & Tel+Instru & Other ID \\
     &  [h:m:sec]       & [$^o$:$'$:$''$]   & [sec] & [sec] &                        \\
\hline
\noalign{\smallskip}
 [GK2001] 8   & 11 05 48.8 & $-$ 76 40 17 &  3050            & 2950    & VLT+MOS        &             \\
 
 ISO-ChaI 126 & 11 08 04.2 & $-$ 77 38 43 &  900             &         & CASLEO+REOSC   & [CCE98] 32   \\
 
 [KG2001] 102 & 11 09 49.3 & $-$ 77 31 20 &  2 $\times$ 3105 &  2960   & VLT+MOS        &              \\
 
 [SGR2003] 1  & 11 09 53.1 & $-$ 77 30 58 &  2 $\times$ 3105 &  2960   & VLT+MOS        &              \\
 
 [GK2001] 30  & 11 09 53.4 & $-$ 77 28 36 &  2 $\times$ 3105 &  2960   & VLT+MOS        & ISO-ChaI 220 \\
 
 Sz 33        & 11 09 53.9 & $-$ 76 29 25 &  3040            &  2960   & VLT+MOS        & ISO-ChaI 224, [KG2001] 110   \\
 
 [GK2001] 31  & 11 09 55.0 & $-$ 76 31 12 &  3040            &  2960   & VLT+MOS        & ISO-ChaI 225     \\
 
 ISO-ChaI 237 & 11 10 11.9 & $-$ 77 35 31 &  900             &         & CASLEO+REOSC   & [CCE98] 48,[OTS99]  45, [KG2001] 121 \\
 
 [GK2001] 53  & 11 14 23.6 & $-$ 77 56 12 &  3300            & 3300    & VLT+MOS        &          \\

\noalign{\smallskip}
\hline
\end{tabular}
 
\vskip 0.2in
 
\noindent
Note: We adopt the SIMBAD source denomination scheme which is based on the
recommendations of the IAU Commission 5 Task Group on Designations.

\end{table}

\begin{table}
\caption[]{Compiled Photometric Data and H$\alpha$ Equivalent Width Measurements} \label{OBS}
\begin{tabular}{lrrrrrrc}
Name & L$^{\mathrm{a}}$ & K$^{\mathrm{b}}$ & H$^{\mathrm{b}}$ &
J$^{\mathrm{b}}$ & I$^{\mathrm{c}}$ & H$\alpha$ (EW[\AA]) & Na I 8183-8195 (EW[\AA]) \\
\hline
\noalign{\smallskip}
 [GK2001] 8       &       & 13.94 & 14.67   & 15.22   &         & $<-$ 5    &  4.9 \\
 
 ISO-ChaI 126     &       &  8.30 & 9.67    & 11.51   &  14.39  & $-$ 97.1  &     \\
 
 [KG2001] 102     & 10.62 & 11.07 & 11.45   & 12.47   &         & $-$ 87.2  &  2.7 \\
 
 
 [SGR2003] 1      &       & 14.87 & 15.64   &         &         & $-$ 182.6 &     \\
 
 
 [GK2001] 30      &       & 12.48 & 13.33   & 14.57   &  18.11  & $-$ 92.3  &  2.1 \\
 
 Sz 33            &  8.22 &  9.10 &  9.84   & 11.19   &  13.99  & $-$ 9.6   &  2.2 \\
 
 [GK2001] 31      &       & 12.75 & 13.73   & 14.84   & 17.12   & $-$ 68.7  &  1.8 \\
 
 ISO-ChaI 237     &  7.69 &  8.55 &  9.34   & 10.75   &  13.90  &  $<-$ 3   &     \\
 
 [GK2001] 53      &       & 13.73 & 14.57   &         &         & $-$ 15.7  &  6.2 \\
\noalign{\smallskip}
\hline
\end{tabular}
 
\vskip 0.2in
 
\begin{list}{}{}
\item[$^{\mathrm{a}}$\citet{kego01}]
\item[$^{\mathrm{b}}$\citet{goke01}]
\item[$^{\mathrm{c}}$\citet{cam98}]
\end{list}
 
\end{table}                                        
\section{Data analysis and results}

\subsection{The sample}

In Table 1 we list our sample stars together with their coordinates
and observing logs; in Table 2 photometric data from the literature
are provided.  Seven of the nine stars in our sample were selected 
among young low mass and brown dwarf candidates proposed by several
recent infrared surveys of the cloud \citep{cam98, per00, goke01, kego01}.
These objects, unlike most of the proposed candidates, have 
optical counterparts 
on the the Digitized 
Sky Survey\footnote{Based on photographic data obtained using The UK Schmidt
Telescope. The UK Schmidt Telescope was operated by the Royal Observatory
Edinburgh, with funding from the UK Science and Engineering Research
Council, until 1988 June, and thereafter by the Anglo-Australian
Observatory. Original plate material is copyright (c) the Royal
Observatory Edinburgh and the Anglo-Australian Observatory. The
plates were processed into the present compressed digital form
with their permission. The Digitized Sky Survey was produced at
the Space Telescope Science Institute under US Government grant NAG W-2166.
Copyright (c) 1993, 1994, Association of Universities of Research in
Astronomy, Inc. All right reserved.} (DSS) plates and thus optical spectra
can be obtained.  The existence of an optical object at 
the position of the infrared discovered source
was the selection criterion used to define the present sample. 

[SGR2003] 1 is the only object
not selected as a candidate very low mass member of the
cloud by previous surveys.  This object was observed
simultaneously with [GK2001] 30 (ISO-ChaI 220). The central slit
of the MOS was placed on [GK2001] 30 whereas the rest of the slits
on optical stars in the field. One of these objects (identified as [SGR2003] 1)
turned out to be a new low mass cloud member judging from the strong
emission in H$\alpha$ (equivalent width $= -$ 182.6). 
Table 1 provides the coordinates of this object and Table 2 K and H
magnitudes \citep{goke01}. [SGR2003] 1 was not detected at J and
thus not placed on the color-color {$\rm (J-H~ vs.~ H-K)$} diagram to 
determine likely near-infrared color excesses. 
Similarly Sz 33, a previously known T Tauri star of the cloud
\citep[see, for example,][]{law96}
lying in the close vicinity of [GK2001] 31 (ISO-ChaI 225),
was observed on the same MOS frame as this object. 

\subsection{The Spectra}

We mention that for five objects in our 
sample (ISO-ChaI 126, [GK2001] 30, Sz 33, [GK2001] 31, and
ISO-ChaI 237) infrared spectra are available
in the literature \citep{gope02,goma03}. However
most of these spectra are almost featureless at the spectral resolution
used, thus not allowing a full characterization of
the objects. [GK2001] 30 shows Br$\gamma$ in emission while Sz 33 
have Pa$\beta$, Pa$\delta$ and Br$\gamma$ in emission. The spectral slope
of these sources, with exception of Sz 33, increases with the wavelength
suggesting the presence of cool material probably in a circumstellar disk. 
Optical low resolution spectra of these objects provide 
additional useful information to
support their pre-main sequence status. In particular 
H$\alpha$ in emission is a distinctive feature of T Tauri
stars and of previously known young BDs. Furthermore, spectral classification
schemes based on optical features are in general better developed
than near-infrared systems and thus can provide more reliable estimates
for the low-mass candidates.

Figs. 1-5 show the spectra of the observed objects.  
Photospheric features such as molecular bands of
CaH, TiO, and VO, as well as atomic lines of Na I and K I 
are indicated. These molecular and atomic features are common among M--type
type stars \citep{kir91}. 

In the following sections we will use the detection of specific
spectral features (such as H$\alpha$ in emission and the Na I
absorption doublet) to unveil the nature of our sample objects. The strength of
the TiO molecular bands at 6180 and 7100 \AA~ and the PC3 index of \cite{mar99} will 
be applied to derive spectral types for these objects.

\subsubsection{Spectral Features}

Table 2 lists H$\alpha$ equivalent width measurements
for the observed objects. We estimate an uncertainty of the order of 
3-5\% due to noise in our spectra and errors in placing the 
continuum level. Five objects in our sample (ISO-ChaI 126, [KG2001] 102, 
[SGR2003] 1, [GK2001] 30 and  [GK2001] 31) show conspicuous H$\alpha$ 
emission line with equivalent widths $>$ 50\AA. Sz 33 and [GK2001] 53 have
moderate emission in this line.  For other two objects
in the sample, [GK2001]8 and ISO-ChaI 237, we do not see a strong
H$\alpha$ in emission and are able to measure only an upper limit to
its equivalent widths. Whereas higher signal to noise and resolution spectra are
required to check in detail for the presence (absence) of H$\alpha$
in emission, these two objects are most likely foreground dwarfs or
background giants.  As a group, spectra in Figs. 1-5 show 
similar characteristics as M--type 
T Tauri stars and previously know young BDs in star forming 
regions \citep[see, for
example,][]{sch77, bri98, com00, mar01}. 
In particular, we note that the equivalent widths of H$\alpha$ are
comparable with those measured by \cite{com00} for their sample of
very low mass stars and BDs in Chamaeleon I.

\citet{mar96} in their study of BDs in the Pleiades found
that, due to lower gravity, the Na I doublet (8183-8195 \AA) tends to be weaker in 
young sub-stellar objects than in field stars of identical spectral type.
These authors analyzed objects with spectral types in the M3.5 -- M9.5 range. 
Table 2 lists the Na I doublet equivalent widths for the objects in our sample for 
which the red part of the spectrum is available and of good quality.
We estimate an uncertainty of
$\sim$ 0.5 \AA~ in our measurements. 
Using the spectral types derived in the next section (see Table 3), we compare the Na I
doublet equivalent widths of our objects with those of Pleiades and field
stars listed by \citet[][-their Table 5]{mar96}.

Three of the objects ([KG2001] 102, [GK2001] 30 and [GK2001] 31)
have very modest Na I doublet equivalent widths, smaller than those
of Pleiades and field stars of the same spectral type,
confirming that they are very low mass young
objects. The Na I doublet equivalent width of [GK2001] 53 is comparable to those of two
Pleiades BDs of the same spectral type, while the equivalent width of [GK2001] 8
is larger than that of the Pleiades BD Calar 4 (with identical spectral type as 
[GK2001] 8) and comparable to
that of field dwarfs. This supports the hypothesis that [GK2001] 8
is not a low mass member of Chamaeleon I and is likely a foreground dwarf.
Finally, Sz 33 with a M2 spectral type is outside the range analyzed
by \citet{mar96}.  

Finally, whereas the Li I $\lambda$ 6707.8 line is
a well-known indicator of youth in late-type fully convective stars 
\citep[see, for example,][]{mag93}, the modest spectral resolution of our spectra, 
in combination with the low S/N ratio in a few cases, prevented
us from detecting this line in our sample objects. 
Higher resolution data are thus required to unambiguously detect this
line in the spectra of the present sample.  

In summary, H$\alpha$ together with the Na I doublet confirm that [KG2001]102,
[GK20001] 30 and [GK2001] 31 are very likely pre-main sequence, low mass
cloud members, while the spectra of [GK2001] 8 and ISO-ChaI 237 do not
support cloud membership; the status of [GK2001] 53 remains uncertain, since
it has an H$\alpha$ comparable to that of previously known low mass
cloud members, but a somewhat high Na I doublet equivalent width.  However
the equivalent width of this doublet for [GK2001] 53 is still comparable to
those of two Pleiades BDs of the same spectral type.  Finally, whereas we
were not able to use the Na I doublet for Sz 33, ISO-ChaI 126, and [SGR2003] 1,
the presence of H$\alpha$ in emission in their spectra supports the pre-main
sequence nature of these sources. In addition Sz 33 is a previously known T Tauri member
of the cloud \citep{law96}. 

\begin{figure}
\centering
\includegraphics[width=16cm]{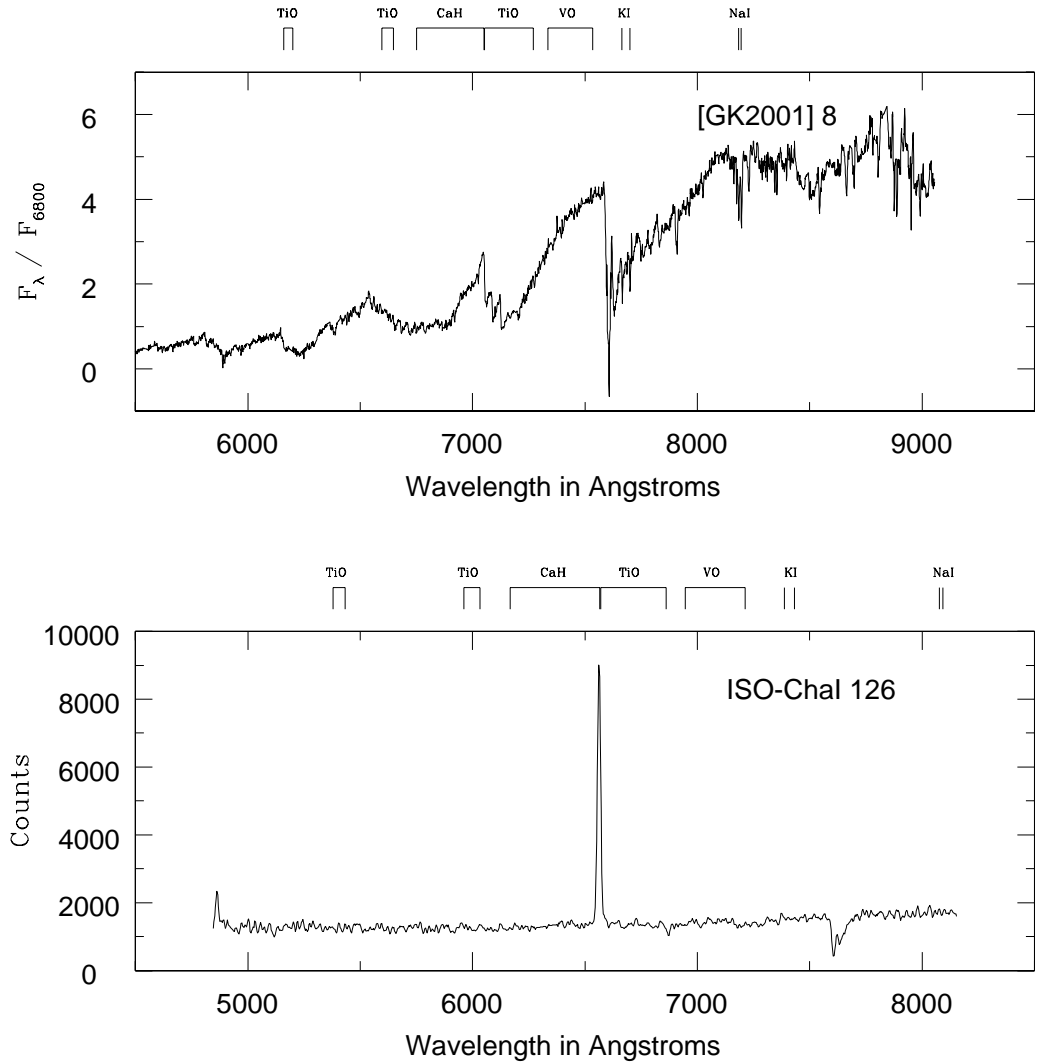}
\caption{Optical spectra of [GK2001] 8 and ISO-ChaI 126 obtained with the
VLT+MOS and the CASLEO+REOSC, respectively. Molecular and atomic features
common in M-type stars are indicated. The [GK2001] 8 spectrum has been 
flux-calibrated and normalized to the flux at \hbox{6800~\AA}.
}
\label{Fig1}
\end{figure}
 
\begin{figure}
\centering
\includegraphics[width=16cm]{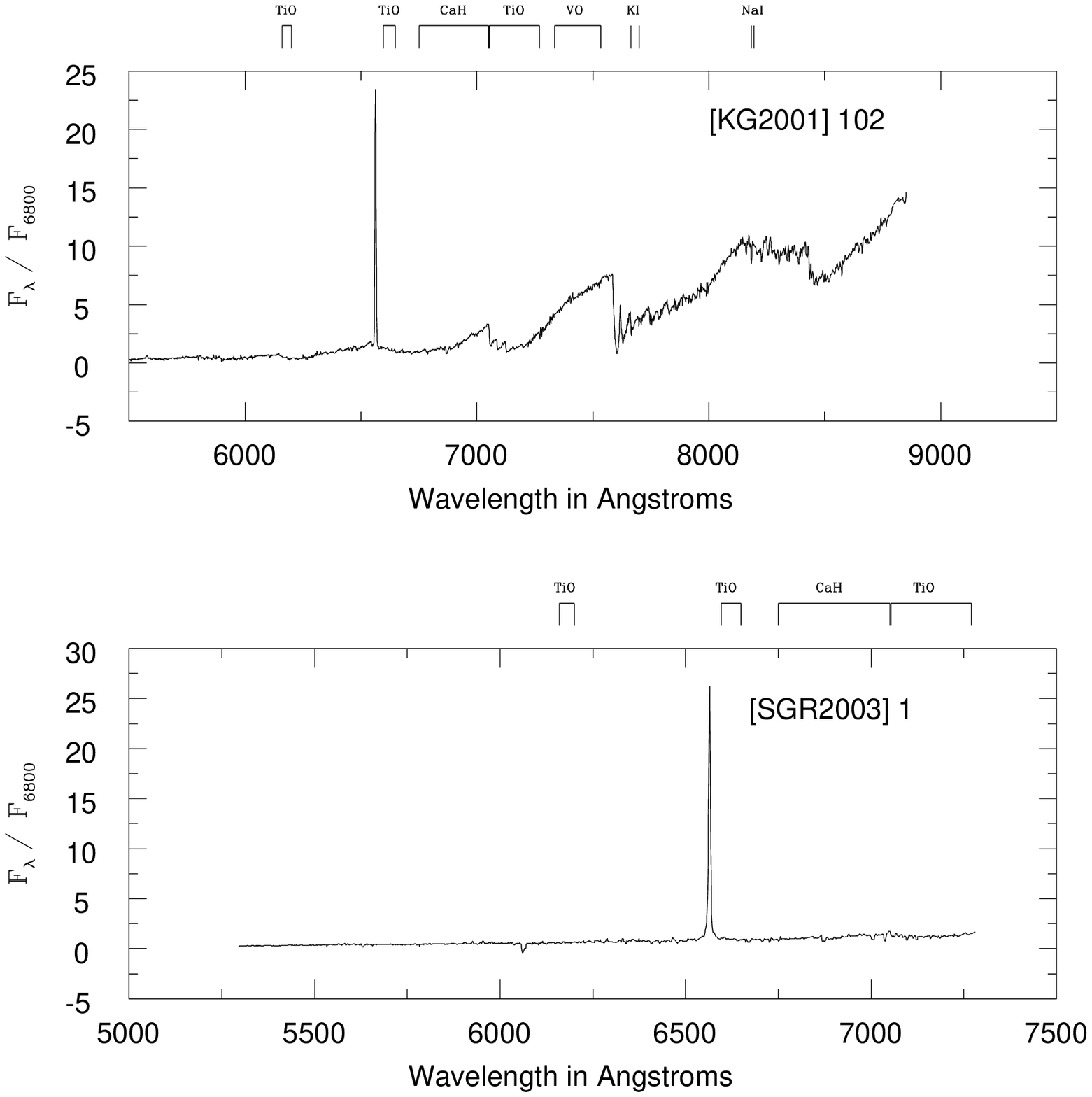}
\caption{Optical spectra of [KS2001] 102 and [SGR2003] 1 obtained with the
VLT+MOS. Molecular and atomic features common in M-type stars are indicated.
The spectra are flux-calibrated and normalized to the flux at 6800 \AA.
}          
\label{Fig2}
\end{figure}
 
\begin{figure}
\centering
\includegraphics[width=16cm]{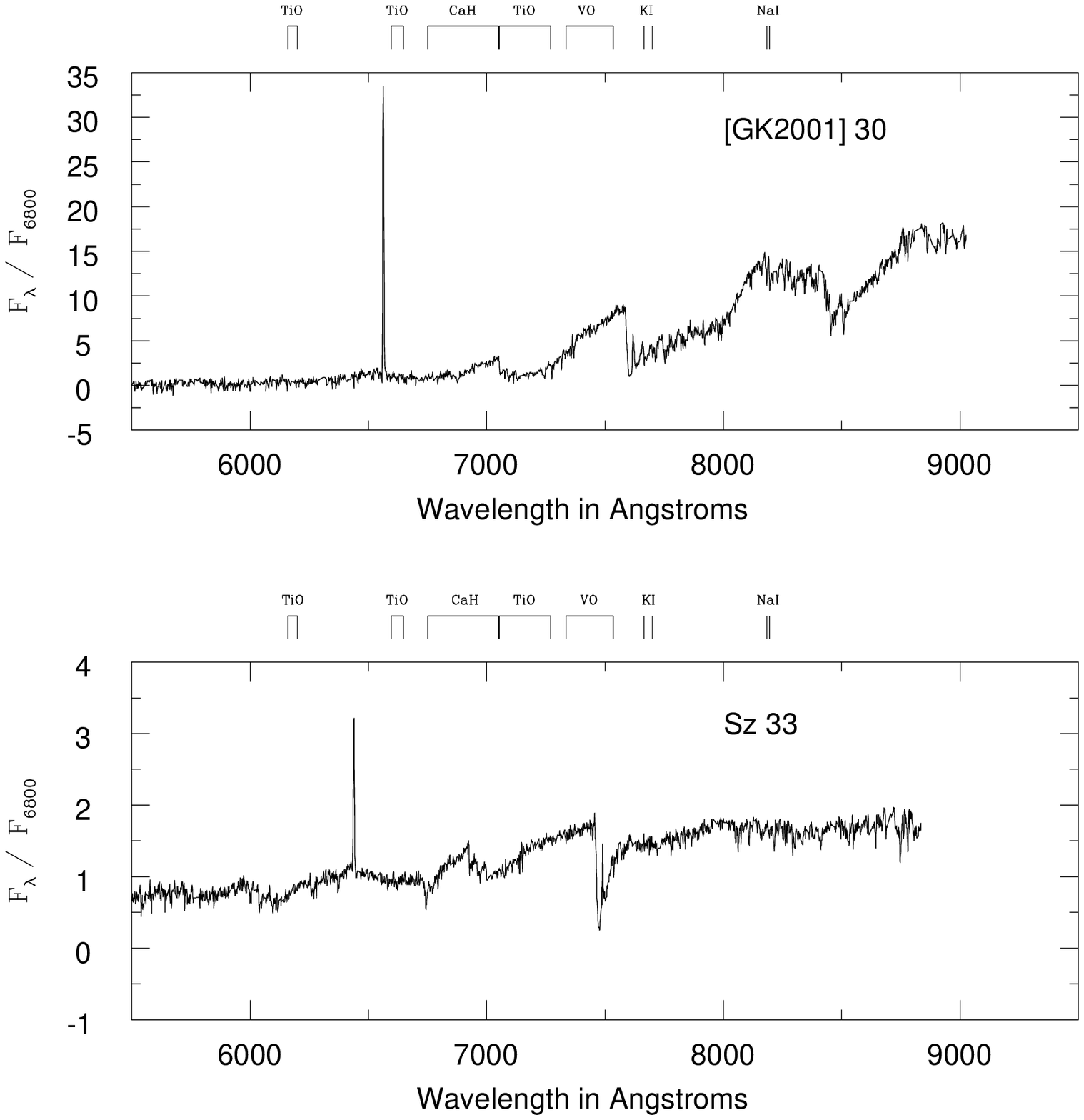}
\caption{Optical spectra of [GK2001] 30 and Sz 33 obtained with the
VLT+MOS. Molecular and atomic features common in M-type stars are indicated.
The spectra are flux-calibrated and normalized to the flux at 6800 \AA.
}          
\label{Fig3}
\end{figure}
 
\begin{figure}
\centering
\includegraphics[width=16cm]{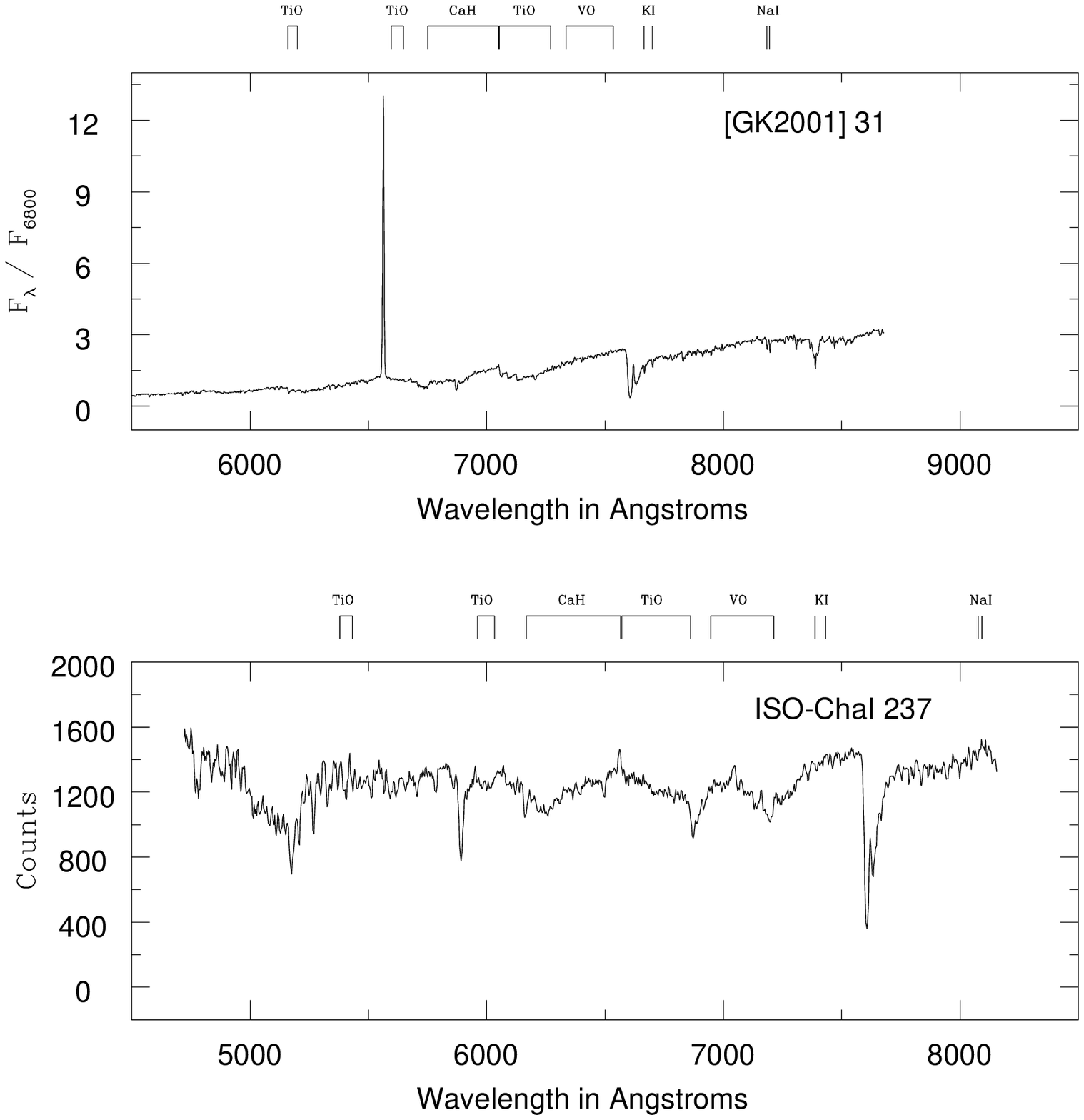}
\caption{Optical spectra of [GK2001] 31 and ISO-ChaI 237
obtained with the VLT+MOS and the CASLEO+REOSC, respectively.
Molecular and atomic features common in M-type stars are indicated.
The [GK2001] 31 spectrum has been
flux-calibrated and normalized to the flux at 6800 \AA.
}          
\label{Fig4}
\end{figure}
 
\begin{figure}
\centering
\includegraphics[width=16cm]{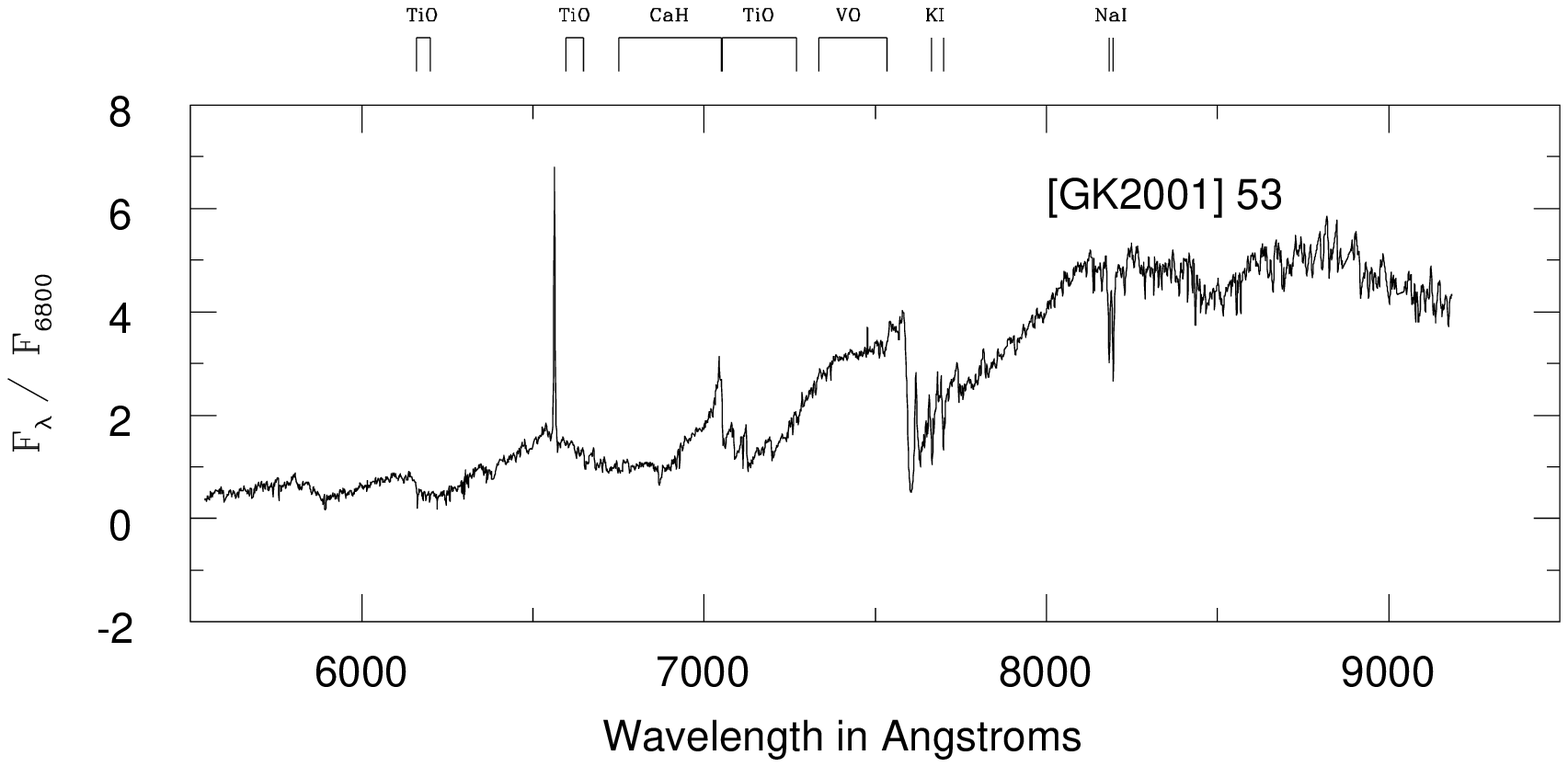}
\caption{Optical spectrum of [GK2001] 53 obtained with the VLT+MOS.
Molecular and atomic features common in M-type stars are indicated.
The spectrum has been flux-calibrated and normalized to the flux
at 6800 \AA.
}          
\label{Fig5}
\end{figure}

\subsubsection{Spectral Types}
                
In order to derive spectral types for our sources,
we used the two TiO absorption
bands at 6180 and 7100 \AA~ that are temperature sensitive for M dwarfs
and giants \citep{oco73}. We also applied the PC3 index form \citet{mar99}. 
For stars with available flux-calibrated FORS spectra, we compared them with the
detailed grid of late type standard stars
obtained by \citet{kir91}, \citet{hen94}, and \citet{kir95}. 

\citet{ken98} applied the Titanium Oxide band indexes at 6180 and 7100 \AA~ to
estimate spectral types for pre-main sequence stars in the Taurus-Auriga
molecular cloud. Each index measures the depth of the TiO band at a given
wavelength relative to an interpolated nearby continuum point. These
authors defined ${\rm [TiO]_1}$ and  ${\rm [TiO]_2}$ as: 

\begin{equation}
{\rm 
[TiO]_1 = - 2.5~ log~ \left[ F_{6180} \over  F_{6125} + 0.225(F_{6370} - F_{6125}) \right]
},
\label{uno}
\end{equation}        

\begin{equation}
{\rm
[TiO]_2 = - 2.5~ log~ \left[ F_{7100} \over  F_{7025} + 0.2(F_{7400} -
F_{7025}) \right]
}
\label{dos}
\end{equation}    

\noindent
\citep[see also][]{oco73}. The 30 \AA~ bandpass filters used in these indexes are
free from strong emission lines (such as H$\alpha$) or telluric
absorptions.  In Fig. 6 we plot these indexes for our sources along with  
the dwarf and giant sequences from \citet{oco73}.  Three objects (ISO-ChaI 126,
[SGR2003] 1 and [GK2001] 30) are not shown in the figure due to the large 
veiling effect that mainly affects the
${\rm [TiO]_1}$ index as we discuss below. 

\begin{figure}
\centering
\includegraphics[width=16cm]{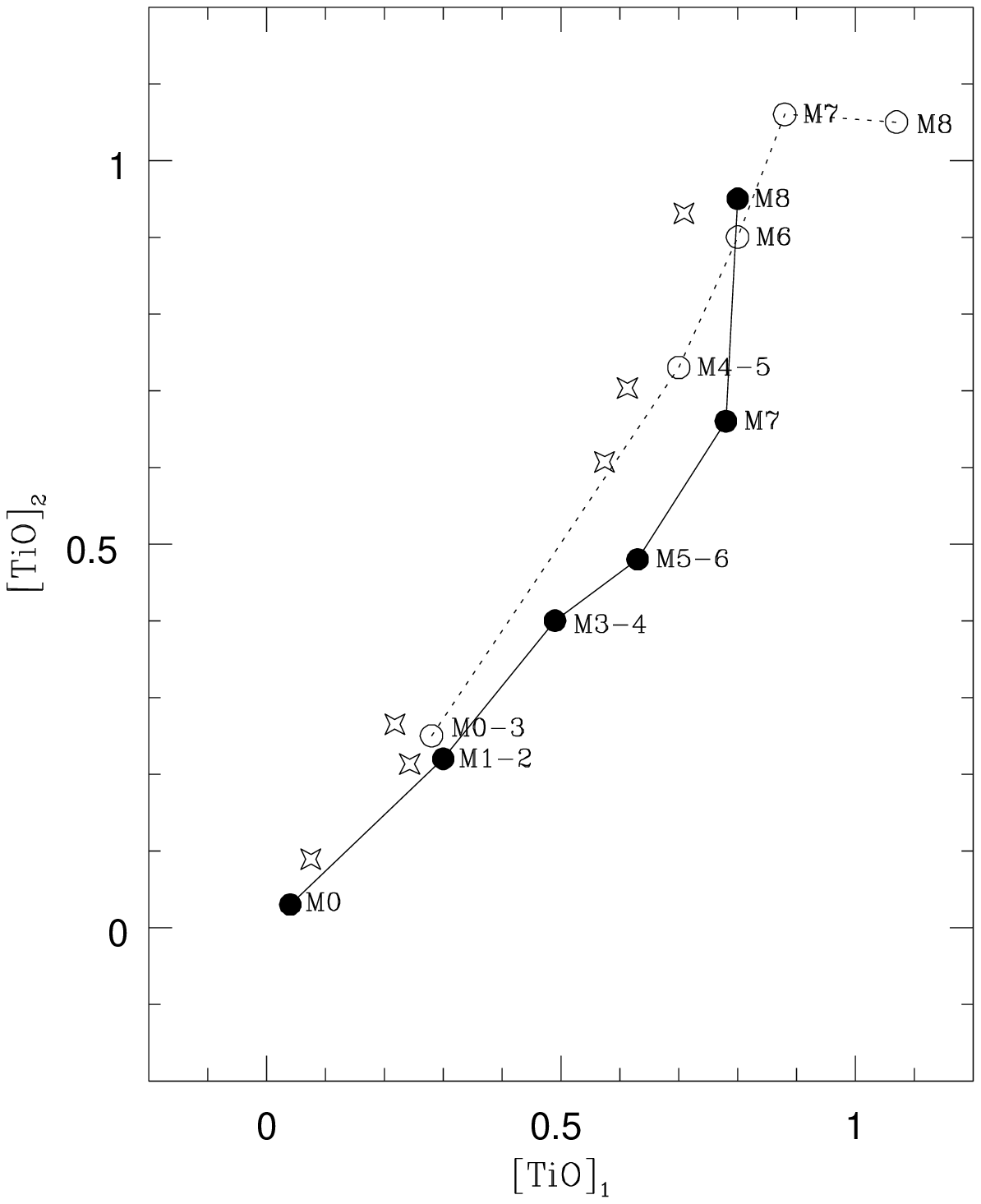}
\caption{${\rm [TiO]_2}$ vs. ${\rm [TiO]_1}$ indexes for field dwarfs
(filled circles) and giants (open circles)
obtained from \citet{oco73}. The continuous line connects the points defining
the dwarf sequence and the dashed line those corresponding to the giant branch. 
Six of the young stellar objects in Tables 1 and 2 are indicated with starred symbols.}
\label{Fig6 }
\end{figure}

The new objects, on average, lie slightly above the sequences defined by the dwarfs and giants, 
having {\it weak} ${\rm [TiO]_1}$ index relatively to the ${\rm [TiO]_2}$
absorption band. \citet{ken98} noticed the same tendency in many T Tauri
stars with strong emission lines in the Taurus cloud. The emission from a hot blue continuum
source \citep[with T $\sim$ 10$^4$;][]{har91} may diminish the depth of the
TiO bands and induce us to assign earlier spectral types to the central star.
This veiling effect increases toward shorter (bluer) wavelengths, affecting
more the band at 6180 \AA~ than the index at 7100 \AA. As mentioned,
five of the stars in
Table 2 have H$\alpha$ equivalent widths $>$ 50 \AA~ which suggests the presence
of veiling in their spectra. 

As to the luminosity class,
pre-main sequence stars are known to have surface gravities intermediate
between those of dwarfs (luminosity class V) and giants (luminosity class III). 
\citet{kir91} identified several features sensitive to the gravity of
M type stars. In particular the intensity of the CaH absorption band at
6975 \AA~ and the Na I lines at 8183-8195 \AA~ decrease with decreasing gravity and
then are stronger in dwarfs than in giants.  They defined
two color ratios (A and C, respectively) that compare the fluxes in a 30 \AA~
band centered on these features with respect to a nearby continuum integrated
over an identical bandpass. These color ratios for our sample give values
intermediate between those of dwarfs and giants and, on average, slightly closer
to the giant branch. We caution that
the veiling effect may also affect the depth of the
gravity sensitive features and hence lead us to derive lower gravities, closer
to class III luminosity objects. 

Three objects need to be discussed in detail.
ISO-ChaI 126 shows a featureless spectrum (except for H$\alpha$ and H$\beta$,
both in emission, see Fig. 1 lower panel); 
for this object we adopted a M2 spectral type derived by
\citet{goma03} from a near-infrared spectrum. 

The R-band spectrum of [SGR2003] 1 has a 
strong H$\alpha$ line in emission and no other spectral feature and
the TiO band at 6180 \AA~ is completely veiled (see Fig. 2 lower panel). 
The I-band region (not shown in Fig. 2 lower panel) was observed under poor sky conditions and 
some night sky lines were registered on the stellar spectrum. However, it 
clearly shows the TiO absorption at 7100 \AA, with a depth corresponding to a
M6 spectral type.  
Due to the veiling effect, it is likely that the depth of this band is
diminished and hence, as discussed,
we are assigning a spectral type earlier than the actual
type. The I-band region also shows the VO band centered at 7445 \AA~ that
appears for spectral types as late as M7 or later \citep{kir95}; 
for the latest
M--subtypes the VO band rather than the TiO band is indeed
a more reliable indicator
of temperature as the TiO starts to saturate \citep{kir91}.
\citet{kir95} defined the VO ratio as: 

\begin{equation}
{\rm VO~ ratio = {0.5625(F_{7350}-F_{7400}) + 0.4375(F_{7510}-F_{7560}) \over F_{7420}-F_{7470}},}
\label{tres}
\end{equation}     

\noindent
where 
{\rm  F$_{\lambda_1}$ $-$ F$_{\lambda_2}$} 
indicates the flux integrated between 
{\rm ${\lambda_1}$ and ${\lambda_2}$}. For M7 stars they obtain 1.02 $\leq$ VO~
ratio $\leq$ 1.07 and 1.07 $\leq$ VO~ ratio $\leq$ 1.12 for M8 stars. We derive
VO~ ratio = 1.08 for [SGR2003] 1 indicating an M7--8 spectral type. 
We adopt an M7 spectral type for this object;
we caution however that we measured the VO band in a
spectrum with sky emission lines superimposed on the stellar continuum.
Although, in first approximation, this has a  
negligible effect on the depth of the detected absorption bands, better quality data
are required to confirm our initial VO ratio measurement. 
We mention that a spectral type earlier than M7 would
shift horizontally to the left this object on the HR diagram (see Fig. 7), 
corresponding to larger values of mass and age. In particular for an M6
spectral type we would obtain an age $>$ 10$^{7}$ yr, which is hard to 
conciliate with the strong H$\alpha$ emission and highly veiled 
spectrum of this object. 
Both features indicate that the object is most likely very young 
and probably not older than a few million years. 
In addition, the comparison of the I-band spectrum of [SGR2003] 1 with the
grid of spectra of \cite{kir91} indicates a spectral type M7 for this source. 

In the case of [GK2001] 30 strong sky lines contaminated the region around the
${\rm [TiO]_1}$ band and prevented us from measuring this index. In addition
the R-band data are featureless (except for the presence of H$\alpha$ in emission)
indicating an increasing veiling effect towards this spectral region. From the depth of
the TiO band at 7100 \AA~ we estimate a spectral type M5 for this object.

\citet{mar99} used the spectral index PC3 to derive spectral types for late-M and
L field dwarfs. This index compares the fluxes in two 40 \AA~ bands centered
at 8250 and 7560 \AA, respectively. For spectral types M2.5 -- L1 they
obtained the following relation:

\begin{equation}
{\rm
SpT = -6.685 + 11.715 \times (PC3) - 2.024 \times (PC3)^2
}
\label{cuatro}
\end{equation}

We calculated this index for our sample objects and used \citet{mar99} calibration 
to estimate spectral types.

Table 3 lists the spectral types base on the TiO-band indexes corresponding
to both the dwarf (left side) and giant (right side) sequences,
as well as those obtained from the index PC3 from \citet{mar99} and those
derived from the comparison with the \citet{kir91}'s
grid.  We also indicate the adopted spectral type for each object. 
For ISO-ChaI 126 the near-infrared spectrum derived spectral type
is given. In the case of [SGR2003] 1 we used the VO band rather than the TiO
band to estimate spectral type as mentioned above. 
We estimate an uncertainty of about $\pm$ 1 subclass for the derived
spectral types based on measurement errors of the TiO and VO bands.
The same uncertainty is expected from spectral types derived by 
the comparison with \citet{kir91}'s grid and from the PC3 index of
\citet{mar99}. The veiling effect may increase this uncertainty up
to $\pm$ 2 for the highly veiled and mostly featureless objects. 

As already mentioned, five of the objects
(ISO-ChaI 126, [GK2001] 30,  Sz 33, [GK2001] 31, and ISO-ChaI 237)
have near-infrared spectra published and derived spectral types and masses
\citep{goma03,gope02}.  
Excluding ISO-ChaI 126 for which we adopted the near-infrared spectral type,
for [GK2001] 30, ISO-ChaI 237, and [GK2001] 31, 
the optical and near-infrared derived spectral
types agree within $\pm$ 1 subtype.  
For Sz 33 we instead estimated a M0 spectral
type from our optical spectrum, while \cite{gope02} obtained a M2 type from the
measured near-infrared water vapor bands. 

In order to derive effective temperatures, intrinsic colors, and 
bolometric corrections for
our targets from the estimated spectral types,
we adopted the calibrations
obtained by \citet{wil99}. 
These relations are valid for spectral types from M2 to M9.

\section{Luminosities, Masses and Ages}

We used the J magnitudes to estimate bolometric luminosities.
This spectral band is less affected by contamination from circumstellar
infrared excess emission than the K or H data. The bolometric
luminosities are calculated from the following expressions:
 
\begin{equation}
{\rm Log (L_{bol} / L_{\sun}) = 1.89 - 0.4 M_{bol}}
\label{cinco}
\end{equation}
 
\begin{equation}
{\rm M_{bol} = J - A_J - DM + BC_J}
\label{seis}
\end{equation}

\begin{equation}
{\rm A_J = 2.63 E(J-H)}
\label{siete}
\end{equation}      
 
\noindent
where DM $=$ 6.0 is the distance modulus \citep[d=160 pc;][]{whi97},
{$\rm {BC_J}$} is the J-band bolometric correction, and {$\rm J$} is the
apparent magnitude.  {$\rm {A_J}$} is the extinction derived from the
\citet{rile85} reddening law, with {$\rm E(J-H) = (J-H) - (J-H)_o$} the color
excess, and {$\rm (J-H)_o$} the intrinsic color. 
For two objects, [SGR2003] 1 and [GK2001] 53,
with no J magnitude measurements we used the K band data in combination
with the \citet{rile85} reddening law, $\rm {A_K = 1.79 E(H-K)}$ (with
${\rm E(H-K) = (H-K) - (H-K)_o}$ and ${\rm (H-K)_o}$ the intrinsic color)
and the K-band bolometric correction,
{$\rm {BC_K}$}, to estimate the bolometric luminosities.

Luminosities for three of the objects (ISO-Cha I 126, [GK2001] 30, and [GK2001]
31) were previously determined by \cite{per00}. Our
determinations agree with those of Persi et al. within a factor of $\sim$ 2, 
with exception of ISO-Cha I 126. For this object,  the main discrepancy comes from the estimated
{$\rm {A_J}$}. \cite{per00} based their determination on the
{$\rm (I-J)$} color obtained by \cite{cam98} while here we use the {$\rm (J-H)$}
color index from \cite{goke01} to compute {$\rm {A_J}$} \citep[cf. Tables 4 and 3 in][]{per00}.  Additional
measurements of the I and H magnitudes may help to clarify this apparent
lack of agreement.  

We chose pre-main sequence evolutionary tracks and isochrones from \cite{dama97}
\footnote{Available at http://www.mporzio.astro.it/$\sim$dantona/prems.html.}
to derive masses and ages for the observed sources.
\citet{goma03} used these models to study the mass and age
distributions of the young stars belonging to the Chamaeleon I dark cloud.
\cite{dama97}'s calculations provide consistent and plausible results
for the group of objects they analyzed in reasonable
agreement with the higher mass members of the cloud \citep[see][]{law96}.                        
Table 3 lists these parameters.  Fig. 7 shows the locations of observed
Chamaeleon I stars on the HR diagram.

\begin{figure}
\centering
\includegraphics[width=16cm]{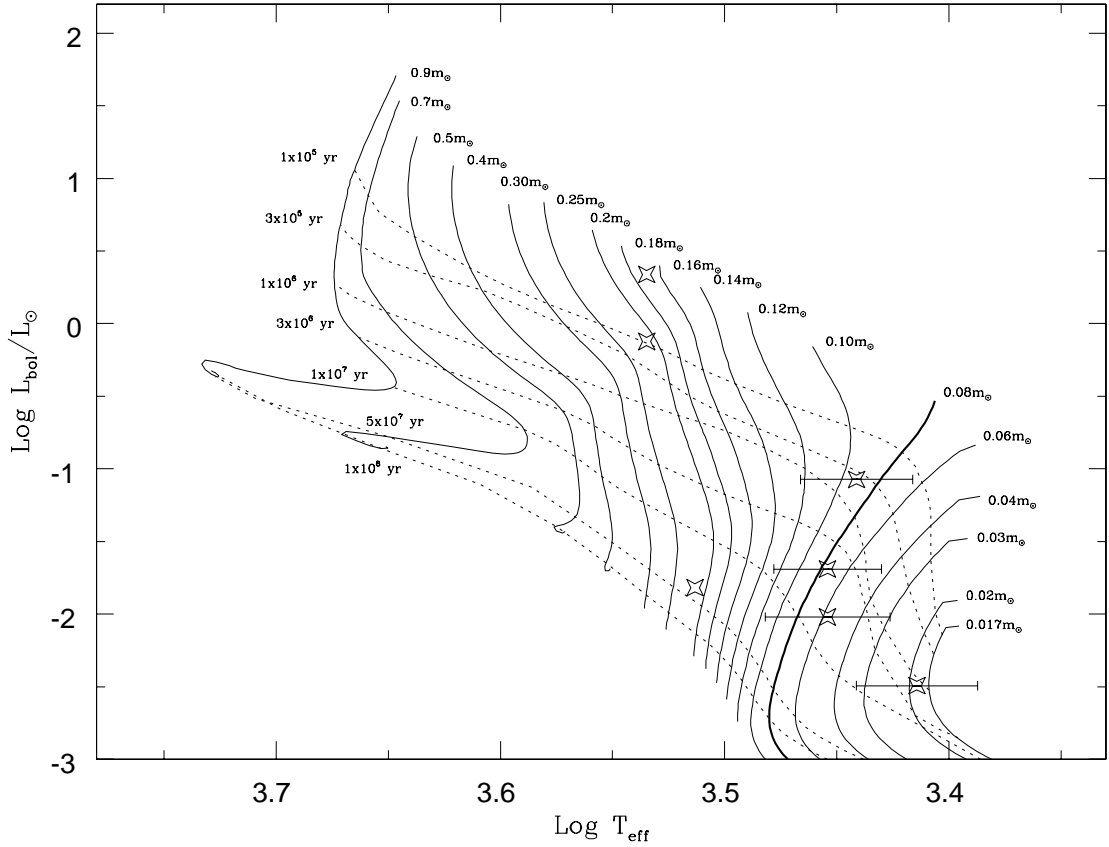}
\caption{HR diagram showing the positions of the seven of the observed stars (starred symbols).
Pre-main sequence evolutionary tracks, indicated with continuous lines,  are
from \cite{dama97}. The thick continuous line corresponds
to 0.08 M\sun~ ($\sim$ 80 M$_{\rm J}$), the H burning limit. The dashed lines correspond to the
isochrones calculated also by the same authors. The horizontal
error bars correspond to a $\pm$ 1 subtype uncertainty in the assigned spectral
types.}
\label{Fig7}
\end{figure}

\begin{table}
\caption[]{Derived Parameters for the Observed Objects} \label{COMP}
\begin{tabular}{lccccclrrr}
Name &  S.T(TiO)$^{\mathrm{a}}$ &  S.T(PC3)$^{\mathrm{b}}$ &
S.T(K91 grid)$^{\mathrm{c}}$ &  S.T(adopted) & T$\mathrm{eff}^{\mathrm{e}}$   & ${\rm {A_J}}^{\mathrm{f}}$ & log ${\rm L_{bol}/L_{\sun}}$ & ${\rm M/M_{\sun}}$ & Age [yr] \\
\hline
\noalign{\smallskip}
 [GK2001] 8   &  M7/M4 &     M5 & M4.5 & M5          &                      &          &                     & & \\

 ISO-ChaI 126 &        &        &      & M2$^{\mathrm{d}}$ &    3426                    & 3.4         &    0.34                      & 0.17  & $<$ 1 $\times$ 10$^{5}$ \\

 [KG2001] 102 &  M8/M6 &     M6 & M6.5 & M6                &    2762                    & 1.1         & $-$1.071                     & 0.09  & 4 $\times$ 10$^{5}$ \\

 [SGR2003] 1  &        &    M7  & M7   & M7                &    2592                    &             & $-$2.49$^{\mathrm{g}}$       & 0.019 & 2 $\times$ 10$^{6}$ \\

 [GK2001] 30  &        &    M6  & M5.5 & M5.5        &    2845                    & 1.7         & $-$1.69                     & 0.075  & 4 $\times$ 10$^{6}$ \\

 Sz 33        &  M2/M2 &        & M2   & M2                &    3426                   & 1.8         & $-$0.12                      & 0.22  & 1 $\times$ 10$^{5}$ \\

 [GK2001] 31  &  M2/M3 &   M3.5 &   M3 & M3                &    3260                   & 1.4         & $-$1.81                      & 0.22  & 4 $\times$ 10$^{7}$ \\

 ISO-ChaI 237 &  M0/M0 &        &      & M0                &        &         &                         &  & \\
 
 [GK2001] 53  &  M6.5/M5 & M5.5 &  M5  & M5.5             &    2845                   &             & $-$2.02$^{\mathrm{g}}$      & 0.06  & 6 $\times$ 10$^{6}$ \\
\noalign{\smallskip}
\hline

\end{tabular}

\begin{list}{}{}
\item[$^{\mathrm{a}}$ Left and right sides from dwarf and giant branches, respectively, \cite{oco73}]
\item[$^{\mathrm{b}}$ Estimated from \cite{mar99}]
\item[$^{\mathrm{c}}$ Estimated from \citet{kir91}'s grid]  
\item[$^{\mathrm{d}}$\citet{goma03}]
\item[$^{\mathrm{e}}$ Obtained from \cite{wil99} calibration]
\item[$^{\mathrm{f}}$ We obtain $\rm {A_K =}$
1.0 and 0.895 for ${\rm [SGR2003]~~1~~and~~[GK2001]~~53}$, respectively]   
\item[$^{\mathrm{g}}$ Derived using K and H magnitudes and
$\rm {A_K = 1.79 E(H-K)}$ obtained from \citet{rile85}]
\end{list}

\end{table}

We excluded from our analysis [GK2001] 8 and ISO-ChaI 237 as they
are probably not young cloud members based on the H$\alpha$ emission
and Na I intensity in the case of [GK2001] 8 (see Sect. 3.2.1.).
Additional data and in particular higher resolution spectra around H$\alpha$ 
are necessary to establish the true nature of these objects. On the other 
hand, we include [GK2001] 53 in this section although the H$\alpha$ and Na I doublet
equivalent width measurements do not allow us to unambiguously assert
(disregard) the pre-main sequence nature of this object.

Four objects in our sample ([KG2001] 102, [GK2001] 30, [SGR2003] 1, and [GK2001] 53) have
masses close or below the H-burning limit (i.e, 0.08 ${\rm M\sun}$, $\sim$ 80 M$_{\rm J}$). 
In Fig. 7, the horizontal error bars indicate
the displacements on the HR diagram that
these objects would experiment due to
$\pm$ 1 sub--type (i.e., $\Delta$T $\sim$ 150 K) uncertainty.
[SGR2003] 1 is a sub-stellar mass object regardless of the spectral
type (or temperature) uncertainty. 
[KG2001] 102, [GK2001] 30, and [GK2001] 53 have masses close to the border
line between the stellar and sub-stellar regimes. 
An uncertainty of $\pm$ 1 sub--type prevent us from
establishing the (stellar or sub-stellar) nature of the objects (see Fig. 7).
The rest of the sources have masses between 0.22 and 0.17 ${\rm M\sun}$. 

Luminosity determinations have a ``typical'' uncertainty of about a factor
of 2 (or $\sim$ 0.03 in log scale). 
However the effect on our mass determinations
is less significant 
than the uncertainty in spectral type or effective temperature
as the evolutionary tracks are almost vertical. 

The masses for the objects for which near-infrared spectra are available
agree within a factor of $\sim$ 2 with those derived from
the spectral classification based on near-infrared spectra, with the differences
in the inferred masses
reflecting the differences in the assigned spectral types. 

Our sample sources span a range of ages between
$\sim$ 10$^5$ yr and few $\times$ 10$^7$ yr, 
in agreement with higher mass members
of the cloud \citep{law96}. In particular, only a small fraction of our
objects (two out of nine) have ages $\geq 10^7$~years, supporting
the conclusions of \cite{com00} and \cite{goma03} on the star formation
rate in the cloud.

For the four objects in Table 3 with masses 
close or below the H-burning limit we compare our
mass and age determinations, based on the \cite{dama97} models, with two
additional sets of pre-main sequence evolutionary tracks and isochrones. 
We select the models of \cite{buw97}\footnote{Available at 
http://zenith.as.arizona.edu/$\sim$burrows/dat-html/data/}
and \cite{cha02}\footnote{Available at ftp://ftp.ens-lyon.fr/pub/users/CRAL/ibaraffe/}
as both include 
detailed calculations below the 0.08 ${\rm M_{\sun}}$ limit. Fig. 8 shows these
tracks and isochrones superimposed on the HR diagram. In Table 4 we
compare masses and ages derived using these models. 
[SGR2003] 1, [GK2001] 30, and [GK2001] 53 have sub-stellar masses, independently
of the adopted calculations, while [KG2001] 102 may be stellar or
sub-stellar depending on the tracks used.

\begin{figure}
\centering
\includegraphics[width=16cm]{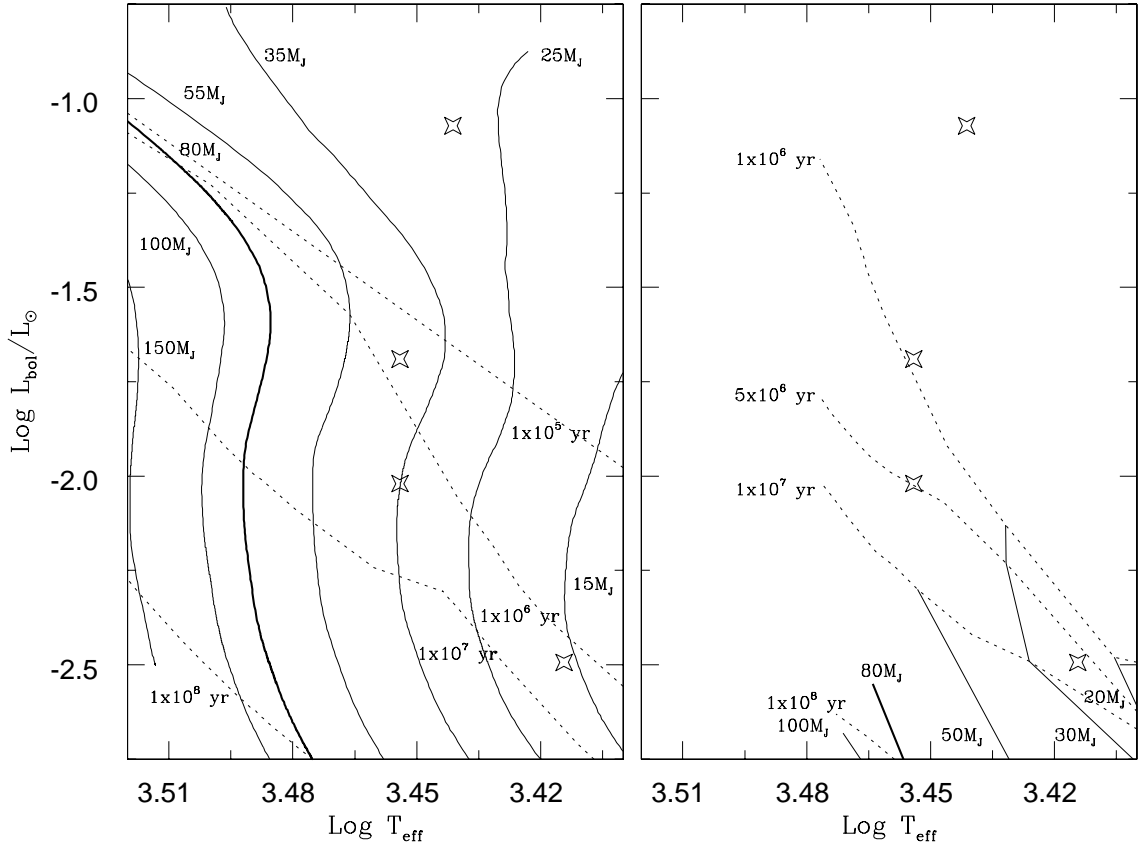}
\caption{Pre-main sequence evolutionary tracks calculated by \cite{buw97}
--left panel-- and \cite{cha02} --right panel--, indicated with continuous lines, 
and superimposed on the HR diagram. The thick continuous line corresponds
to 0.08 M\sun~ ($\sim$ 80 M$_{\rm J}$), the H burning limit. The dashed
lines show the isochrones calculated also by the same authors. The positions of four of the
observed stars (see Table 4) are indicated with starred symbols.}
\label{Fig8}
\end{figure} 

\begin{table}
\caption[]{Comparison of Masses and Ages Corresponding to Different Models} \label{DERV}
\begin{tabular}{l|lr|lr|lr}
Name & ${\rm M/M_{\sun}}$  & Age [yrs] & ${\rm M/M_{\sun}}$  & Age [yrs] & ${\rm M/M_{\sun}}$       & Age [yrs] \\
     &  B97$^{\mathrm{a}}$ &     &  C02$^{\mathrm{b}}$ &      & DM97$^{\mathrm{c}}$               &       \\       
\hline
\noalign{\smallskip}
 [KG2001] 102 &  0.027    & $<$ 10$^{5}$         &         &                     & 0.09   & 4 $\times$ 10$^{5}$  \\    
 
 [SGR2003] 1  &  0.016    & 5 $\times$ 10$^{6}$  & 0.025   & 7 $\times$ 10$^{6}$ & 0.019  & 2 $\times$ 10$^{6}$ \\
 
 [GK2001] 30  &  0.045    & 7 $\times$ 10$^{5}$  & 0.04    & 1 $\times$ 10$^{6}$ & 0.075  & 4 $\times$ 10$^{6}$ \\
  
 [GK2001] 53  &  0.035    & 4 $\times$ 10$^{6}$  & 0.04    & 5 $\times$ 10$^{6}$ & 0.06   & 6 $\times$ 10$^{6}$ \\
\noalign{\smallskip}
\hline

\end{tabular}                            
\vskip 0.2in
 
\begin{list}{}{}
\item[$^{\mathrm{a}}$\citet{buw97}]
\item[$^{\mathrm{b}}$\citet{cha02}]
\item[$^{\mathrm{c}}$\cite{dama97}]
\end{list}                                        

\end{table}

\section{Summary and conclusions}

We present optical spectra of eight candidate very low mass members
and a previously known T Tauri star of the Chamaeleon I dark cloud.
All but two of these objects have H$\alpha$ in emission with equivalent widths
similar to other known very low mass T Tauri stars and previously identified
young BDs \citep{law96, bri98, neco99, mar01}, indicating
their young age.  The Na I doublet (8183-8195 \AA) is weaker
than in stellar objects of the same spectral type for [KG2001] 102, [GK2001]
30, and [GK2001] 31, supporting the very low mass nature of these objects
\citep{mar96}. In the case of [GK2001] 53 the Na I doublet strength is comparable
to two known BDs in the Pleiades. 
Higher resolution spectra of
these objects would allow the detection of the Li I $\lambda$ 6707.8 and thus
an independent confirmation of their youth. 

The lack of significant H$\alpha$ emission in ISO-ChaI 237 and [GK2001] 8
casts doubts on the nature of these objects. The 
Na I doublet strength for [GK2001] 8 suggests that this object is 
probably a field dwarf.  Higher resolution spectra are required to
properly classify both ISO-ChaI 237 and [GK2001] 8.

In summary, out of eight candidates, two turned out not to be cloud members, 
the nature of other object ([GK2001] 53) would require additional confirmation
by higher resolution spectroscopy, while H$\alpha$, together with the Na I 
doublet strength in some cases, support the pre-main sequence nature and low
mass classification of the remaining five sources. 

We used the strength of the TiO or VO molecular bands
as well as the PC3 index from \citet{mar99} and 
the grid of spectra of \cite{kir91}, to derive spectral types and adopted \cite{dama97}
evolutionary tracks and
isochrones to estimate masses and ages. We detect one new BD and
three transition stellar/sub-stellar objects of the cloud. \cite{buw97} 
and \cite{cha02} models confirm these results.
The rest of the observed objects have masses between 0.22 and 0.17 ${\rm M\sun}$.
[SGR2003] 1, with a derived mass of 0.019 ${\rm M\sun}$ ($\sim$ 19 M$_{\rm J}$),
lies very close to the deuterium burning limit ($\sim$ 13 M$_{\rm J}$).
This is the least massive object identified in the cloud so far.
These sources span a range of ages between $\sim$ 10$^5$ yr and few $\times$
10$^7$ yr, in agreement with higher mass members of the cloud \citep{law96}.
 
The Chamaeleon I dark cloud has been searched at different
wavelengths to detect sub-stellar mass objects during the last few years
\citep{cam98, oas99, per00, goke01}. \cite{com00} reported the detection of
four bona-fide BDs in the Chamaeleon I dark cloud in addition to eight candidate 
transition stellar/sub stellar objects \citep[see also,][]{neco99,neu02}.
\cite{gope02} and \cite{goma03} identified $\sim$ 30 very low mass objects,
including several transition objects.  These investigations have significantly
increased the low mass population of the cloud \citep[see][]{goma03}.
However, these authors attempted a determination of the IMF of
the cloud and noticed an apparent scarcity of very
low mass objects in the whole Chamaeleon I cloud based on the behavior of the
IMF in the central 300 arcmin$^2$ region, determined by \citet{com00}.
This small portion of the cloud has been intensively surveyed at different
wavelengths combining both photometric and spectroscopic information.
\cite{goma03} estimated that about 100 objects remain to be identified in the
whole cloud.  

The results presented in this paper contribute to the census of the
pre-main sequence population of the cloud. In particular,
we identified one sub-stellar
object, increasing the number of bona-fide BDs in the cloud from 4 to 5. The
newly discovered BD as well as the three transition objects 
analyzed in this contribution will eventually help to determine
the behavior of the IMF for the complete cloud in the sub-stellar regimen and
a comparison with other nearby star-forming clouds.

The newly detected objects increase the overall number of young BDs and
transition objects known today. These objects will eventually allow us to
carry out statistical studies of the physical properties of young BDs. 
To better understand similarities and differences among stars, brown dwarfs
and planets, confrontations of the characteristics of the three groups are
needed. 
                                                                                         
\begin{acknowledgements}
We are grateful to the ESO staff for assistance
with the observations, specially to Martino 
Romaniello for help during the VLT Phase II preparation. 
We are grateful to
Dr. Eduardo L. Mart\'\i n, the referee,
for helpful criticism that improved the
content and presentation of this paper.    
This research has made use of the SIMBAD database,
operated at CDS, Strasbourg, France. The CCD and data
acquisition system at CASLEO has been partly financed
by R. M. Rich trough U.S. NSF grant AST-90-15827.

\end{acknowledgements}

\end{document}